# The Metallicity Dependence of the CO-to-H$_2$ Conversion Factor from Observations of Local Group Galaxies

Christine D. Wilson

Department of Physics and Astronomy, McMaster University, Hamilton Ontario Canada

L8S 4M1

– 2 –

## ABSTRACT

Six additional molecular clouds have been mapped in the nearby metal-poor dwarf irregular galaxy IC 10 using the Owens Valley Millimeter-Wave Interferometer. Since a good distance measurement is now available for IC 10, it is possible to measure accurately the CO-to-$H_2$ conversion factor in this galaxy. The result for IC 10 is combined with published data for four other Local Group galaxies (M31, M33, NGC 6822, and the SMC) to trace the dependence of the CO-to-$H_2$ conversion factor on the oxygen abundance using a well-defined sample of extragalactic giant molecular clouds. These data show conclusively that the CO-to-$H_2$ conversion factor increases as the metallicity of the host galaxy decreases, with the conversion factor increasing by a factor of 4.6 for a factor of 10 decrease in metallicity. This new calibration of the effect of metallicity on the CO-to-$H_2$ conversion factor will allow more accurate molecular gas masses to be obtained in galaxies and regions of galaxies with moderately low metallicity.

*Subject headings:* ISM: Molecules – Galaxies: Local Group – Galaxies: ISM – Galaxies: Individual (IC 10, M31, M33, NGC 6822, SMC)

## 1. Introduction

Accurate measurements of the molecular gas content of galaxies are important for determining total gas content and star formation efficiency and are a critical input parameter to models of galactic chemical evolution, interstellar medium structure, and star formation laws. Molecular gas masses are commonly determined from the CO J=1-0 luminosity via a conversion factor between CO intensity and $H_2$ column density. However, the accuracy of these gas masses depends on how well we understand the dependence of this conversion factor on the metallicity, density, and temperature of the clouds (Dickman, Snell, & Schloerb 1986). In particular, the metallicity dependence of the CO-to-$H_2$ conversion factor is quite controversial. Some calculations show that the CO emission from an ensemble of clouds in a distant galaxy does not depend strongly on metallicity as long as the individual clouds are optically thick in CO (Dickman et al. 1986). However, models of individual low-metallicity clouds suggest that such clouds are optically thin in CO, which produces a strong dependence of the CO-to-$H_2$ conversion factor on metallicity (Maloney & Black 1988). If the CO-to-$H_2$ conversion factor is strongly dependent on metallicity, then molecular gas masses in irregular galaxies, the outer regions of spiral galaxies, and other low metallicity systems may be significantly underestimated.

Giant molecular clouds in our own Galaxy have diameters in the range of 10 to 100 pc (Sanders, Scoville, & Solomon 1985). By obtaining *resolved* observations of individual giant molecular clouds, we can determine directly the CO-to-$H_2$ conversion factor for each galaxy by comparing the mass calculated using the virial theorem with the observed CO luminosity. Such observations are currently possible only for galaxies in the Local Group, since sufficient sensitivity and linear resolution cannot be obtained with existing millimeter-wave interferometers for galaxies beyond 1-2 Mpc. Fortunately, Local Group galaxies cover a range of metallicity of at least a factor of 10 and so provide us with a good opportunity to study the effect of metallicity on the CO-to-$H_2$ conversion factor. In this paper we present new observations of giant molecular clouds in the Local Group irregular galaxy IC 10, for which a good distance determination has recently become available (Wilson et al. 1995). We combine these results with published observations from four other Local Group galaxies (M31, M33, NGC 6822, and the SMC), for which good distances, oxygen abundances, and resolved observations of molecular clouds are available in the literature, to determine the dependence of the CO-to-$H_2$ conversion factor on metallicity over the range solar to one-tenth solar.

## 2. New CO Observations of IC 10

Two overlapping fields in IC 10 centered on Region 3/5 (Becker 1990) were observed in the CO J=1-0 line with the Owens Valley Millimeter-Wave Interferometer between 1991 April 12 and June 6. The field centers $(\alpha(1950), \delta(1950))$ were $(00^h17^m38.8^s, +59^o04'33.0'')$ and $(00^h17^m38.8^s, +59^o03'48.0'')$. The data were obtained using the 1 MHz (2.6 km s$^{-1}$) 32 channel filterbank, and typical double-sideband system temperatures were 200-400 K. Absolute flux calibration was obtained by observations of Uranus and is estimated to be accurate to 30%. Cleaned maps were made using the routine MXMAP in the AIPS software package for each of the 1 MHz channels individually. The rms noise in an individual channel map was 0.15 Jy beam$^{-1}$ and the synthesized beam is $5.5'' \times 9.5''$ or $22 \times 38$ pc at a distance of 0.82 Mpc (Wilson et al. 1995). This distance has been obtained from near-infrared observations of Cepheid variables and is in good agreement with the somewhat more uncertain distance of 0.95 Mpc obtained from the luminosities of the Wolf-Rayet and blue stars in IC 10 (Massey & Armandroff 1995). The data for the two fields were combined into a single mosaic cube including weighting for the primary beam, so that more accurate measurements could be obtained for any emission in the region where the two fields overlapped. The integrated intensity map is shown in Figure 1.

Clouds were identified and their properties measured from the mosaic channel maps using the procedures described in Wilson & Scoville (1990). The molecular mass $M_{mol}$ is given by

$$M_{mol} = 1.61 \times 10^4 d_{Mpc}^2 S_{CO} \; M_\odot \qquad (1)$$

(Wilson & Scoville 1990) where $S_{CO}$ is the CO flux in Jy km s$^{-1}$ and $d_{Mpc}$ is the distance in megaparsecs. This formula assumes the Galactic value for the CO-to-H$_2$ conversion factor, $\alpha_{Gal} = 3 \times 10^{20}$ cm$^{-2}$ (K km s$^{-1}$)$^{-1}$ (Strong et al. 1988; Scoville & Sanders 1987) and includes a factor of 1.36 to account for the presence of helium. Molecular masses were calculated for all clouds, while virial masses were only calculated for resolved clouds. Clouds were considered resolved if their deconvolved diameters in both directions were at least two-thirds of the beam size. For the virial mass, we assume a $1/r$ density profile, which gives

$$M_{vir} = 99 V_{FWHM}^2 D_{pc} \; M_\odot \qquad (2)$$

where $V_{FWHM}$ is the full width half maximum velocity in km s$^{-1}$ and $D_{pc} = 1.4 \overline{D}_{FWHM} = 0.7(D_\alpha + D_\delta)$ is the cloud diameter in parsecs (see Wilson & Scoville 1990 for details). The properties of the six new clouds identified here, as well as the three clouds from Wilson & Reid (1991) corrected to the new distance, are given in Table 1. Note that the velocity width of MC3 was incorrectly measured in Wilson & Reid (1991); the correct value is given in Table 1.

## 3. The Metallicity Dependence of the CO-to-H$_2$ Conversion Factor

To map out the metallicity dependence of the CO-to-H$_2$ conversion factor we combine these new data for IC 10 with published data for four other Local Group galaxies: M31,

M33, NGC 6822, and the SMC. All five galaxies have accurate distance determinations from observations of Cepheid variables, either from near-infrared data, for which the reddening is both less severe and easily measured (IC 10, NGC 6822, SMC), or from four-color optical data, from which both the reddening and the distance can be obtained (M31, M33) (Table 2). The oxygen abundances in these galaxies are also known from many observations of HII regions, and for the two spiral galaxies the radial gradient in the oxygen abundance is known. Finally, the detailed properties of individual giant molecular clouds are available in the literature, either from interferometric observations (M31, M33, NGC 6822, IC 10) or from high-resolution (15 pc) single dish observations (SMC). We do not include the LMC, since high-resolution data for individual clouds are not available in the literature. Although the Milky Way also has an abundance gradient, we do not include any Galactic clouds in our sample, since distances to individual Galactic clouds are much more uncertain than for extragalactic molecular clouds and the derived CO-to-$H_2$ conversion factor depends inversely on the distance.

For the interferometric data, the virial and molecular gas masses were calculated using equations (1) and (2) and the distances given in Table 2. For the SMC data, deconvolved full-width half-maximum diameters are not published, and so in equation (2) we use the diameter as given in Rubio, Lequeux, & Boulanger (1993), rather than 1.4 times the full-width half-maximum diameter. Calculations suggest that the virial masses calculated in this way may be slightly larger compared to the method used in the other four galaxies. The CO-to-$H_2$ conversion factor for each cloud can then be obtained from

$$\alpha = \alpha_{Gal} \frac{M_{vir}}{M_{mol}} \qquad (3)$$

For the two spiral galaxies, oxygen abundances at the radius of each cloud were estimated from the known abundance gradient (M31: Zaritsky, Kennicutt, & Huchra 1994; M33: Vilchez et al. 1988). The oxygen gradient in M31 is shallow enough that molecular clouds at 8 and 12 kpc have very similar oxygen abundances, and so one average conversion factor was calculated using all the M31 clouds. However in M33, the abundance gradient is quite steep in the inner disk (Vilchez et al. 1988, Zaritsky 1992) and so the conversion factor was calculated for three different subsets of the data: the nucleus (MC2-4, Wilson & Scoville 1990), the inner disk region (MC6, MC10, MC20, MC27, MC31-32, Wilson & Scoville 1990), and the two giant HII regions (NGC 604-1, NGC 604-2, NGC 604-4, and NGC 595-1, Wilson & Scoville 1992).

The value of the CO-to-$H_2$ conversion factor is plotted versus the oxygen abundance in Figure 2 and shows a small but significant slope. The uncertainties in the values of $\alpha/\alpha_{Gal}$ given in Table 2 are the statistical uncertainties in the mean of the individual cloud samples; however, the systematic uncertainties are likely to be larger. In particular the systematic uncertainty in the absolute calibration of the CO fluxes is estimated to be $\sim 20\%$, and it is this uncertainty which is plotted in Figure 2 and used in the least-squares fit. A least-squares fit to the data excluding the upper limit for NGC 6822 gives the relation

$$\alpha/\alpha_{Gal} = (5.95 \pm 0.86) - (0.67 \pm 0.10)[12 + log(O/H)] \qquad (4)$$

This equation gives $\alpha = \alpha_{Gal}$ for $12 + log(O/H) = 8.88$ and the fit has $\chi^2 \sim 7$. The slope obtained here corresponds to a factor of 4.7 increase in the CO-to-$H_2$ conversion factor for a factor of 10 decrease in the metallicity. *Thus data for individual molecular clouds in five Local Group galaxies show conclusively that the CO-to-$H_2$ conversion factor increases as the metallicity of the host galaxy decreases.*

A similar analysis of the metallicity dependence of the CO-to-$H_2$ conversion factor is presented in Arimoto, Sofue, & Tsujimoto (1995). Although the two analyses agree on the value of the CO-to-$H_2$ at $12 + log(O/H) = 8.90$ to within 3%, the Arimoto et al. study obtains a slope of -0.80, rather than the shallower slope of -0.67 obtained here. Although the two slopes agree at the 1-2$\sigma$ level, the Arimoto et al. sample includes data for M51, which is much too distant to allow individual clouds to be resolved, as well as low spatial resolution data (150-200 pc) for the SMC (Rubio et al. 1991) and the LMC (Cohen et al. 1998). Inspection of Figure 2 in Arimoto et al. suggests that the slope is being steepened by the M51 data at the high-abundance end and by the low-spatial resolution data for the SMC and the LMC at the low-abundance end. We would argue that these data should not be included in an analysis of the CO-to-$H_2$ conversion factor because there is no guarantee that the large structures being observed are actually gravitationally bound, and thus that the galaxies used in this paper represent a better selected sample.

Care must be taken in applying this calibration of the CO-to-$H_2$ conversion factor to galaxies with unusual star formation properties, such as starburst and ultraluminous galaxies. An enhanced ultraviolet radiation field may increase the temperature of the clouds and produce a smaller conversion factor than the one obtained from considering the variation in metallicity alone (Maloney & Black 1988). However, if the giant molecular clouds in these galaxies are also denser than clouds in normal galaxies, the effects of changing density and temperature may partially cancel one another (Scoville & Sanders 1987). With the exception of the clouds near the giant HII regions in M33, the clouds in this sample are not found in extreme star-forming environments, and thus the possible effect of the ultraviolet radiation field on the CO-to-$H_2$ conversion factor has been neglected in this study. Observations of molecular clouds in the disk of M33 with oxygen abundances similar to those of the two giant HII regions would be helpful in assessing the uncertainty in the metallicity calibration of the conversion factor due to variations in the local ultraviolet radiation field.

This new calibration of the effect of metallicity on the CO-to-$H_2$ conversion factor will allow more accurate molecular gas masses to be obtained in galaxies with moderately low metallicity, such as irregular galaxies and galaxies at moderate redshift, and in spiral galaxies with known abundance gradients. However, an extrapolation of this relation to abundances outside the range probed by the Local Group should be used with caution, since the physical mechanisms which produce the observed relation are not well understood. A similar empirical determination of the effects of density and temperature on the CO-to-$H_2$ conversion factor must await the next generation of millimeter interferometers, with which resolved observations of giant molecular clouds in the nearest starburst galaxies will be possible.

## 4. Conclusions

Six additional molecular clouds have been mapped in the nearby metal-poor dwarf irregular galaxy IC 10 using the Owens Valley Millimeter-Wave Interferometer. We have combined the data for IC 10 with published data for molecular clouds in four other Local Group galaxies (M31, M33, NGC 6822, and the SMC) to trace the dependence of the CO-to-$H_2$ conversion factor on the oxygen abundance using resolved observations of extragalactic giant molecular clouds. These data show conclusively that the CO-to-$H_2$ conversion factor increases as the metallicity of the host galaxy decreases, with the conversion factor increasing by a factor of 4.7 for a factor of 10 decrease in metallicity. This new calibration of the effect of metallicity on the CO-to-$H_2$ conversion factor will allow more accurate molecular gas masses to be obtained in moderately low metallicity regions, such as irregular galaxies, the outskirts of spiral galaxies, and galaxies at moderate redshifts.

C. D. W. was partially supported by NSERC Canada through a Women's Faculty Award and Research Grant.

Table 1: Properties of Molecular Clouds in IC 10 (wide plano table, separate LaTeX file)

| Galaxy | Distance | $\alpha/\alpha_{Gal}$ [a] | $12 + log(O/H)$[a] | References | Comments |
|---|---|---|---|---|---|
| M31 | 0.77 Mpc | $1.07 \pm 0.16$ | $8.98 \pm 0.04$ | 1,6,7,13 | 3 clouds |
| M33 | 0.84 Mpc | $0.49 \pm 0.11$ | $9.02 \pm 0.16$ | 2,8,14 | 3 clouds, nucleus |
|  |  | $1.2 \pm 0.1$ | $8.89 \pm 0.03$ | 2,8,14 | 6 clouds, inner disk |
|  |  | $1.5 \pm 0.3$ | $8.48 \pm 0.04$ | 2,9,14 | 4 clouds, giant HII regions |
| IC 10 | 0.82 Mpc | $2.7 \pm 0.5$ | $8.16 \pm 0.12$ | 3,10,15 | 5 clouds |
| NGC 6822 | 0.49 Mpc | $< 2.2 \pm 0.8$ | $8.20 \pm 0.05$ | 4,11,15,16,17 | 3 clouds |
| SMC | 61 kpc | $4.2 \pm 0.4$ | $7.98 \pm 0.02$ | 5,12,18 | 11 clouds |

Table 2: The CO-to-$H_2$ Conversion Factor in Local Group Galaxies

---

[a]The quoted uncertainty is the statistical uncertainty in the mean $(\sigma/\sqrt{N})$, rather than the dispersion of the set of measurements.

References (Distances): 1. Freedman & Madore 1990. 2. Freedman, Wilson, & Madore 1991. 3. Wilson et al. 1995. 4. McAlary et al. 1983. 5. Welch et al. 1987.

References (Molecular Clouds): 6. Vogel, Boulanger, & Ball 1987. 7. Wilson & Rudolph 1993. 8. Wilson & Scoville 1990. 9. Wilson & Scoville 1992. 10. this paper. 11. Wilson 1994. 12. Rubio et al. 1993.

References (Abundances): 13. Zaritsky et al. 1994. 14. Vilchez et al. 1988. 15. Lequeux et al. 1979. 16. Pagel et al. 1980. 17. Skillman et al. 1989. 18. Pagel et al. 1978.

---

TABLE 1
Properties of Molecular Clouds in IC 10

| Cloud | $\alpha(1950)$ ($^h$ $^m$ $^s$) | $\delta(1950)$ ($^\circ$ $'$ $''$) | $V_{peak}$ (km s$^{-1}$) | $V_{FWHM}{}^a$ (km s$^{-1}$) | $T_B$ (K) | $S_{CO}{}^a$ (Jy km s$^{-1}$) | $D_\alpha \times D_\delta{}^{abc}$ (pc) | $M_{vir}{}^{ac}$ ($10^5$ M$_\odot$) | $M_{mol}{}^{ad}$ ($10^5$ M$_\odot$) |
|---|---|---|---|---|---|---|---|---|---|
| MC1 | 00 17 44.3 | 59 00 25 | -331 | 10.9 | 2.9 | 19.6 | $28 \times 42$ | 5.8 | 2.1 |
| MC2 | 00 17 45.0 | 59 00 19 | -324 | 8.9 | 2.6 | 11.7 | $\leq 20 \times 29$ | $\leq 2.7$ | 1.3 |
| MC3 | 00 17 44.0 | 59 00 17 | -331 | 12.8 | 2.1 | 12.9 | $\leq 17 \times 38$ | $\leq 6.3$ | 1.4 |
| MC4 | 00 17 36.9 | 59 04 09 | -336 | 8.0 | 1.3 | 8.9 | $\leq 20 \times \leq 30$ | $\leq 2.2$ | 0.96 |
| MC5 | 00 17 39.1 | 59 03 25 | -329 | 6.8 | 1.7 | 8.9 | $\leq 20 \times 41$ | $\leq 2.0$ | 0.96 |
| MC6 | 00 17 39.4 | 59 04 42 | -329 | 5.3 | 1.5 | 6.0 | $< 15 \times 41$ | ... | 0.65 |
| MC7 | 00 17 39.5 | 59 04 28 | -331 | 3.8 | 2.1 | 3.7 | $27 \times < 25$ | ... | 0.40 |
| MC8 | 00 17 40.9 | 59 04 41 | -334 | 4.7 | 1.3 | 5.9 | $29 \times < 25$ | ... | 0.64 |
| MC9 | 00 17 38.6 | 59 04 54 | -344 | 6.5 | 1.1 | 2.8 | $\leq 18 \times < 25$ | ... | 0.30 |

A distance to IC 10 of 0.82 Mpc is assumed throughout

[a]The typical errors are: $D_{\alpha,\delta} \pm 7$ pc, $V_{FWHM} \pm 1.3$ km s$^{-1}$, $S_{CO} \pm 30\%$, $M_{mol} \pm 30\%$, $M_{vir} \pm 40\%$

[b]$D_\alpha$ and $D_\delta$ are the deconvolved full-width half-maximum diameters in the right ascension and declination directions.

[c]Diameters between 2/3 and 1 times the beam size are indicated by the $\leq$ sign, although clouds are considered resolved if their diameters are at least 2/3 the beam size.

[d]$M_{mol}$ is calculated using the Galactic CO-to-H$_2$ conversion factor (see text)

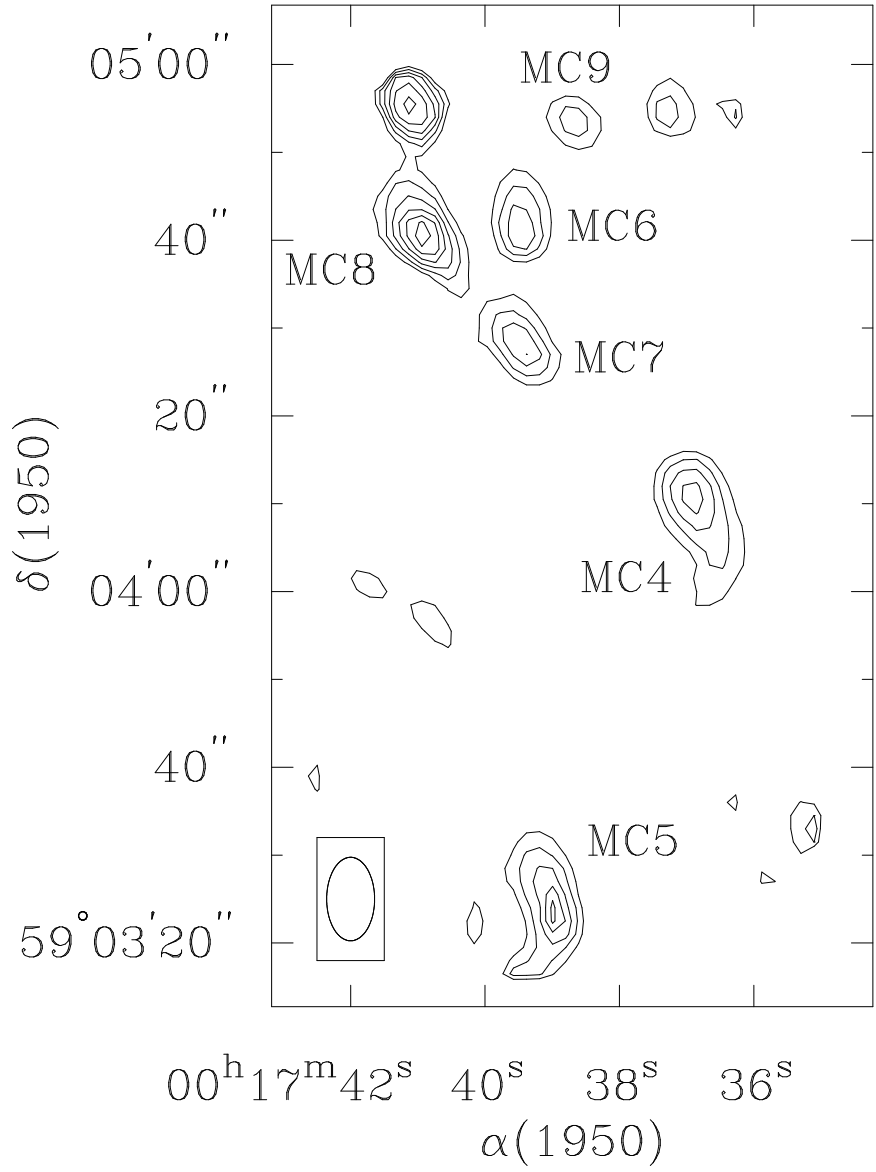

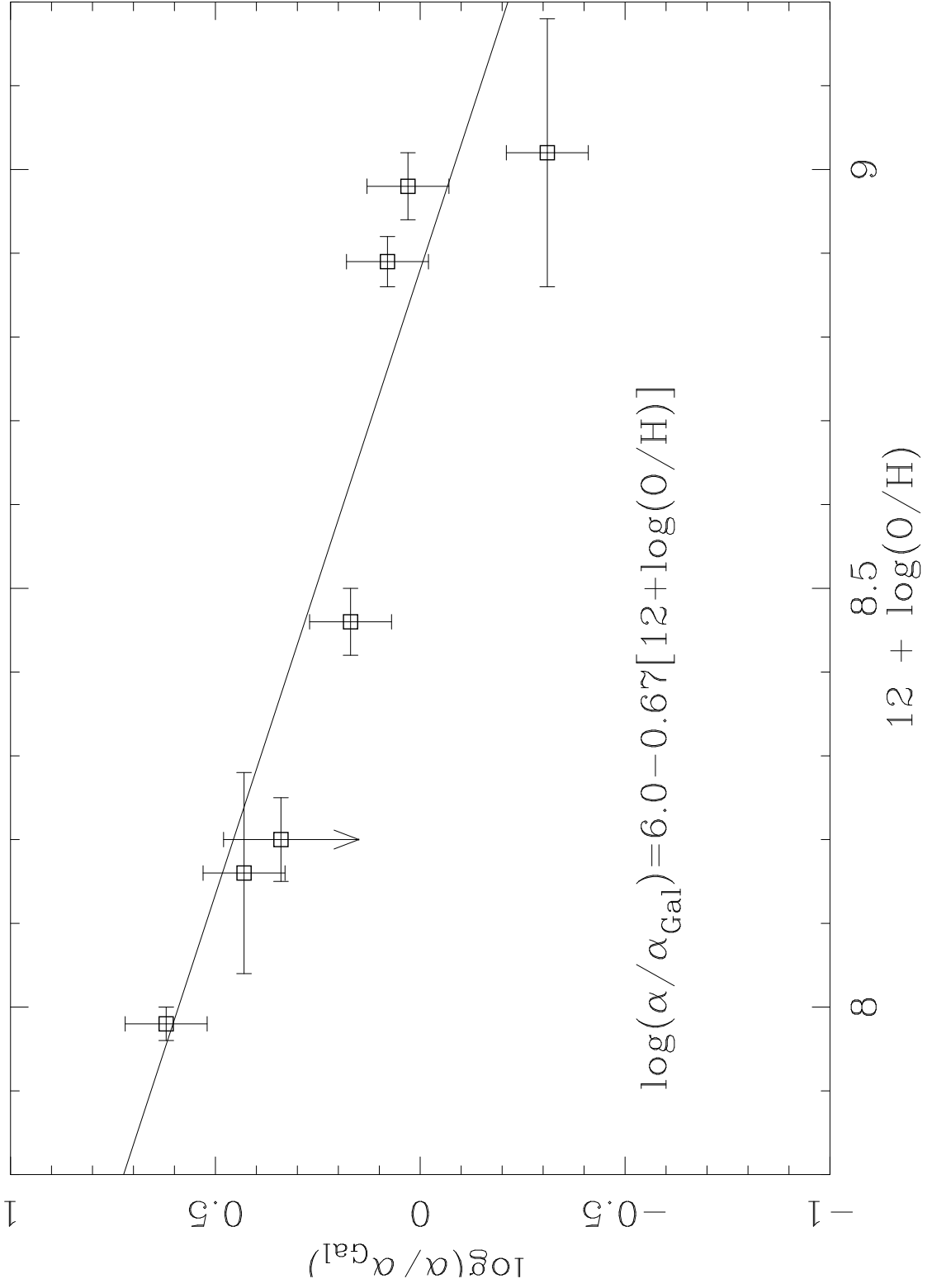

Fig. 1.— The integrated intensity map of the emission in IC 10. The first contour level is 2.2 Jy beam$^{-1}$ km s$^{-1}$ ($2\sigma$) and the contours increase by 1.1 Jy beam$^{-1}$ km s$^{-1}$. The six clouds discussed in the text are labeled. The unlabeled features are noise; the feature in the north-west corner appears strong because of the primary beam correction. The synthesized beam (FWHM) is shown in the lower left corner.

Fig. 2.— The average CO-to-H$_2$ conversion factor in five Local Group galaxies (M31, M33, IC 10, NGC 6822, SMC) is plotted as a function of the oxygen abundance. The data for M33 are broken down into three radial bins, while the measurement for NGC 6822 is an upper limit. The line is a least-squares fit to the data, excluding the NGC 6822 data point.